\documentstyle[aps,twocolumn,prl,psfig]{revtex}

\begin{document}
\draft

\title{\Large Neutron Scattering Studies on Magnetic Structure
 of the Double-Layered Manganite La$_{2-2x}$Sr$_{1+2x}$Mn$_2$O$_7$
($0.30 \leq x \leq 0.50$)}

\author{M. Kubota,$^{\dagger}$ H. Fujioka,$^{\ddagger}$ K. Ohoyama$^{3}$
K. Hirota,$^{\ddagger}$ Y. Moritomo,$^{4}$ H. Yoshizawa,$^{\dagger}$ and Y.
Endoh$^{\ddagger}$}

\address{$^{\dagger}$Neutron Scattering Laboratory, I.S.S.P., University of
Tokyo, Tokai, Ibaraki, 319-1106, Japan}
\address{$^{\ddagger}$CREST, Department of Physics, Tohoku University, Aoba-ku,
Sendai, 980-8578, Japan}
\address{$^{3}$Institute for Materials Research Tohoku University, Aoba-ku,
Sendai 980-8577, Japan}
\address{$^{4}$Center for Integrated Research in Science and Engineering, Nagoya
University, Nagoya, 464-01, Japan}

%\maketitle

\twocolumn[\hsize\textwidth\columnwidth\hsize\csname @twocolumnfalse\endcsname
\maketitle

\begin{abstract}
Systematic powder diffraction studies have been carried out to establish the
magnetic phase diagram of La$_{2-2x}$Sr$_{1+2x}$Mn$_2$O$_7$ (LSMO327) in a wide
hall concentration range ($0.30 \leq x \leq 0.50$), using the HERMES 
diffractometer.  
LSMO327 exhibits a planar ferromagnetic structure for $0.32\leq
x\leq 0.38$ at low temperatures. A finite canting angle between planar magnetic
moments on neighboring planes starts appearing around $x\sim 0.40$ and
reaches 180$^{\circ}$ (A-type antiferromagnet) at $x=0.48$.  At
$x=0.30$, on the other hand, the magnetic moments are aligned parallel to the
$c$-axis.
\end{abstract}

\vspace{0.5cm}
{\it Keywords:} A: magnetic materials, B: crystal growth, C: neutron
scattering, D: magnetic structure
\vspace{0.5cm}

]

%\newpage
%\section{Introduction}

%\begin{center}
%1.\hspace{1zw}{\bf INTRODUCTION}
%\end{center}

Perovskite Mn oxide R$_{1-x}$A$_{x}$MnO$_{3}$ (R: Rare-earth ion, A:
Alkaline-earth 
element) provides an ideal stage to systematically study complex physics
resulting from spin-charge-lattice degrees of freedom.\cite{Urushi95,Millis95} 
Recently, Moritomo {\it et al.}\cite{Moritomo95,Moritomo96} have discovered
that the layered Perovskite Mn oxide La$_{2-2x}$Sr$_{1+2x}$Mn$_2$O$_7$ (LSMO327)
with $x=0.40$ shows colossal magnetoresistance (CMR), which is much more
enhanced than that of similarly doped La$_{1-x}$Sr$_x$MnO$_{3}$.  LSMO327
($I/4mmm$, $a=3.871$~\AA\ and $c=20.126$~\AA.\cite{Mitchell97,K.Hirota98}) has
MnO$_2$ double layers separated by (La$_{1-x}$Sr$_x$)$_2$O$_2$, stacking along
the $c$-axis.

The magnetic structure of LSMO327 has been established between $x=0.40$ and
$0.48$ by neutron-diffraction study using single crystals.\cite{K.Hirota98}  The
low-temperature magnetic phase consists of planar ferromagnetic (FM) and A-type
antiferromagnetic (AFM) components, indicating a canted AFM ordering.  With
increasing $x$, the canting angle between planes changes from 6.3$^{\circ}$ at
$x=0.40$ to 180$^{\circ}$ at $x=0.48$, while the FM ordering temperature $T_C$
decreases from 120~K to 0~K.  It was also discovered that the A-type AFM
ordering remains above $T_C$ up to $T_{N}\sim 200$~K.  At $x=0.30$, on the other
hand, it is reported that the low-temperature magnetic structure is AFM with the
easy axis {\em parallel} to the $c$-axis and that the easy-axis shows a canting
at higher temperature, giving the
in-plane components.\cite{T.Kimura97,T.G.Perring98}  As for
$x=0.50$, superlattice peaks are observed by electron diffraction\cite{J.Q.Li98}
as well as neutron diffraction,\cite{M.Kubota98} which are ascribed to
charge-ordering.  The A-type AFM structure is realized
at low temperatures, while additional magnetic reflections are seen at
intermediate temperatures, indicating the coexistence of CE-type and A-type
AFM structures.

LSMO327 exhibits a variety of magnetic structures with changing temperature
and the hole concentration $x$.  There have been, however, no systematic study
of the magnetism of LSMO327 in a wide temperature and hole concentration range
so far.  To establish the magnetic phase diagram, we have carried out extensive
neutron-diffraction studies on LSMO327 powder samples prepared in consistent
manner.

%\section{Experimental}
%\begin{center}
%2.\hspace{1zw}{\bf EXPERIMENTAL}
%\end{center}

The prescribed amount of dried La$_2$O$_3$, SrCO$_3$, and Mn$_3$O$_4$ are
thoroughly mixed and calcined in the air at 1200--1450$^{\circ}$C for 4 days 
with frequent grinding.  Sample rods were melt-grown in a floating zone optical
image furnace, then powderized again.  All the samples were characterized by
x-ray diffractions, which show no detectable impurities.

We have taken powder diffraction patterns at room temperature,
intermediate temperature ($100-120$~K), and low temperature ($\sim 10$~K)
using HERMES,\cite{K.Ohoyama}  which is a powder neutron
diffractometer with multi-detectors with the Ge $(3\ 3\ 1)$ monochromater
($\lambda=1.819$~\AA).  Temperature dependences of magnetic
reflections were measured using the triple-axis spectrometers GPTAS and TOPAN, 
where the $(0\ 0\ 2)$ reflection of pyrolytic graphite (PG) was
used to monochromate, together with PG filters to eliminate higher order
contaminations.  These spectrometers are located in the JRR-3M reactor in JAERI.

%\section{Results and Discussions}

%\begin{center}
%3.\hspace{1zw}{\bf RESULTS AND DISCUSSION}
%\end{center}

Figure~1 shows five possible models of magnetic structure and
corresponding scattering patterns for LSMO327.  Here we assume that spins are
ferromagnetically coupled within a plane, i.e., FM or A-type AFM phase.  Provided
that spins are antiferromagnetically modulated within the $ab$ plane, there
should be magnetic superlattice peaks at $(h\ k\ l)$ with half integer of $h$
and/or $k$, which is clearly not the case for the present study.  Models a 
(FM-I) and b (FM-II) correspond to FM structures with spins aligning within the
$ab$ plane and along the $c$-axis, respectively.  Models c (AMF-I), d (AFM-II)
and e (AFM-III) represent A-type AFM structures with different alternating
pattern along the $c$-axis.  Note that a canted AFM magnetic structure is
represented by a combination of FM and AFM components, which thus gives a
combination of FM and AFM reflection patters shown in Fig.~1.

\begin{figure}[htb]
\centering

\psfig{file=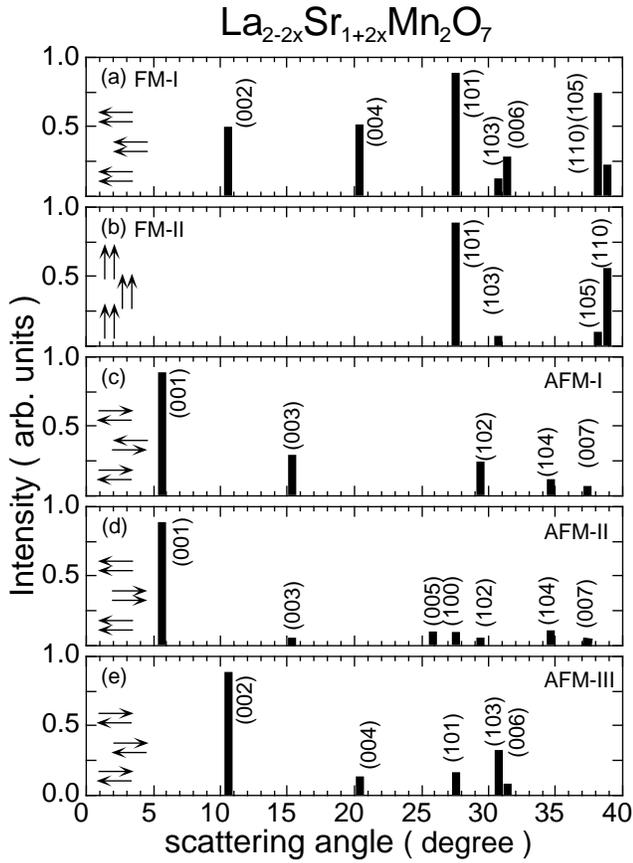,width=8.5cm}
\caption{Five possible models and corresponding reflections for magnetic
structures which would explain the present results of
La$_{2-2x}$Sr$_{1+2x}$Mn$_{2}$O$_{7}$ ($0.30 \leq x \leq 0.50)$.}
\label{Fig:Models}
\end{figure}

\begin{figure}[htb]
\centering

\psfig{file=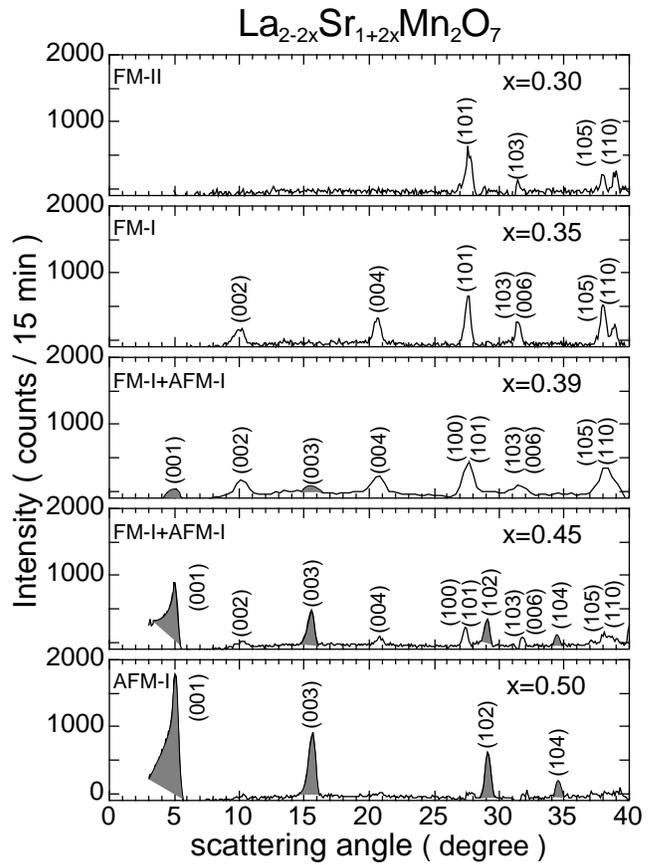,width=8.5cm}
\caption{Powder patterns in the low temperature phase ($\sim 10$~K) of
La$_{2-2x}$Sr$_{1+2x}$Mn$_{2}$O$_{7}$ ($0.30 \leq x \leq 0.50$).  The data
taken at room temperature is subtracted from each profile as background. 
Hatched peaks correspond to AFM reflections, and the others correspond to
FM reflections.}
\label{Fig:Patterns}
\end{figure}

We have carried out comprehensive powder neutron-diffraction measurements at
$x=0.30$, 0.32, 0.335, 0.35, 0.38, 0.39, 0.40, 0.45, 0.48 and 0.50. 
Figure~\ref{Fig:Patterns} summarizes the results for $x=0.30$, 0.35, 0.39,
0.45 and 0.50 at 10~K.  We have noticed that the powder diffraction patterns
for $x=0.32\leq x\leq 0.50$ at low temperatures can be explained by a
combination of FM-I and AFM-I phases: the FM-I phase is dominant at low doping
region, while the AFM-I phase (hatched peaks) gradually takes over the FM-I
phase
with increasing $x$.  The present result is consistent with
Ref.~\onlinecite{K.Hirota98} and makes it clear that the planar FM phase is
stable below $x \sim 0.40$.  We also found that the diffraction pattern at
$x=0.30$ is completely different from the other concentrations, and corresponds
to the FM-II phase, in which spins are aligned ferromagnetically along the
$c$-axis.  The drastic change in the magnetic structure indicates that there
exists a compositional phase boundary around $x=0.30-0.32$.
To study the intermediate temperature phase, we have also meausured powder
diffractions around $100-120$~K.  As is consistent with
Ref.~\onlinecite{K.Hirota98}, only the AFM-I phase remains for $x=0.39-0.50$,
while no magnetic reflections are observed for $x=0.32-0.38$.  As $x=0.30$,
however, the low-temperature FM-II phase disappears and the AFM-II phase appears
at intermediate temperatures.

\begin{figure}[htb]
\centering

\psfig{file=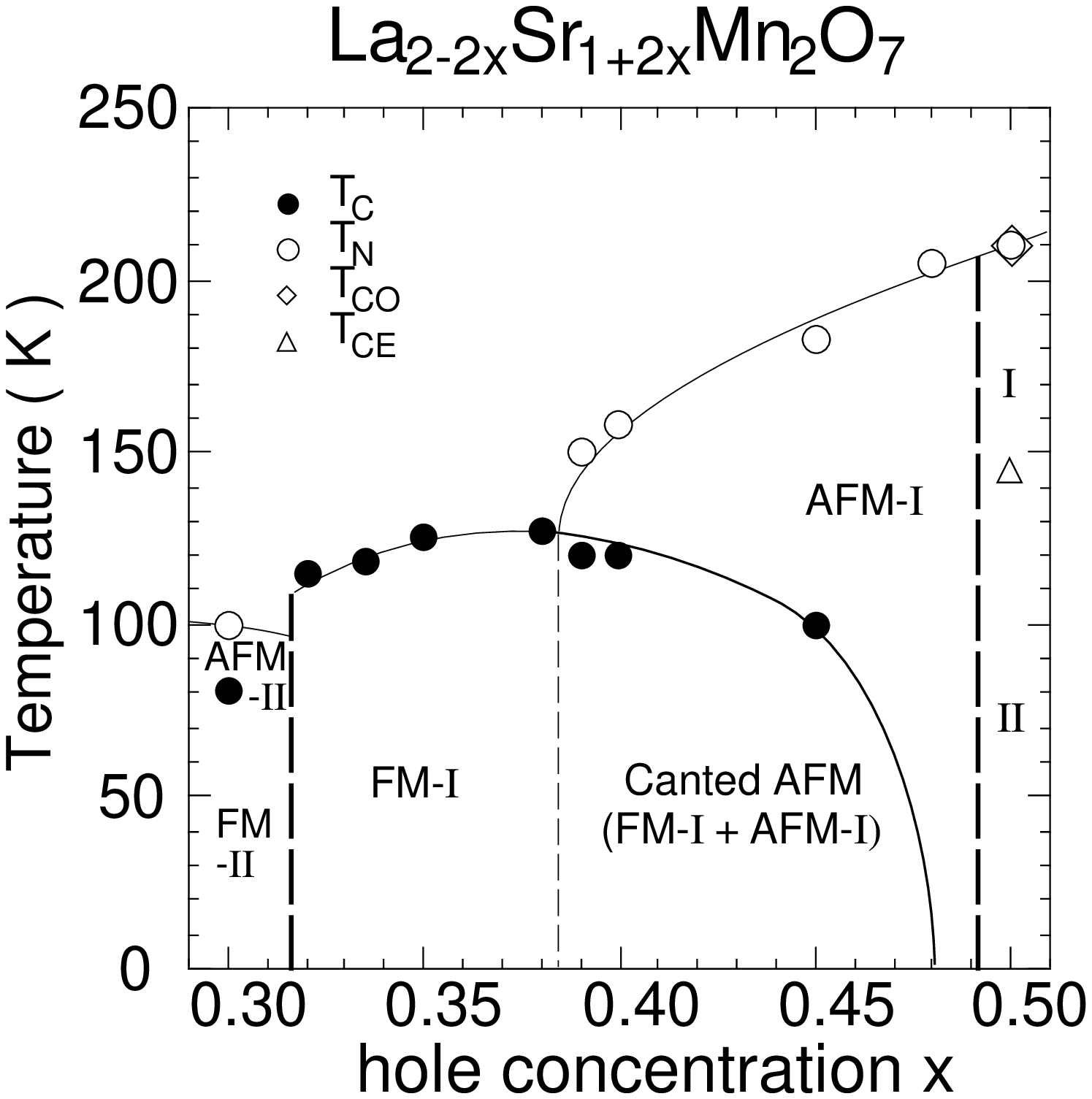,width=7.5cm}
\caption{Magnetic phase diagram of La$_{2-2x}$Sr$_{1+2x}$Mn$_{2}$O$_{7}$ ($0.30
\leq x \leq 0.50$). AFM-I (AFM-II) indicates the planar A-type AFM structure
with FM (AFM) intra-bilayer coupling and FM (AFM) inter-bilayer coupling. 
FM-I and FM-II stand for the FM structures with spin within the $ab$ plane and
along the $c$-axis, respectively. (See Fig.~\protect\ref{Fig:Models})  As for
$x=0.50$, only AFM-I exists in the phase I, while AFM-I and
CE-type AFM coexist in the phase II.}
\label{Fig:Phase}
\end{figure}

We also measured the temperature dependence of typical magnetic
reflections to complete the magnetic phase diagram of LSMO327, as shown in
Fig.~\ref{Fig:Phase}.  As for $0.32\leq x\leq 0.48$, there is essentially one
low-temperature magnetic phase, i.e., a planar canted AFM, in which the canting
angle is zero below $x=0.38$,  becomes finite at $x=0.39$ then gradually
increases with increasing $x$, and finally reaches 180$^{\circ}$ at $x=0.48$. 
Note that the intermediated AFM-I phase appears at $x=0.39$, where the canting
angle in the low-temperature phase becomes finite.  The magnetic phase becomes
more complicated at $x=0.50$ because of the charge
ordering.\cite{J.Q.Li98,M.Kubota98}  Since superlattice peaks due to the charge
ordering and CE-type AFM ordering  are not confirmed
in the present powder diffraction study because those peaks are very weak, we
refer to the results of Ref.~\onlinecite{M.Kubota98} in our phase diagram.

The magnetic phases for $x=0.30$ are completely different from the other
concentrations.  With decreasing temperature, it becomes the planar AFM-II phase
at $\sim 100$~K and then changes to the uniaxial FM-II phase at 70~K.  Perring
{\it et al.}\cite{T.G.Perring98} also conclude that the low-temperature phase is
AFM with spins aligning along the $c$-axis, which is inconsistent with the FM-II
structure in the present study, though the change of the easy-axis direction
from planar to uniaxial at intermediate temperature is similar.  Note that there
exists a compositional phase boundary at $x=0.30-0.32$, which is consistent with
difficulties in growing a crystal.  More detailed study, particularly with
a good
single crystal, is required to completely clarify the magnetic structure around
$x=0.30$.

Recent studies indicate that there is a close relation between the
magnetism and structure of LSMO327, particularly through the Mn $e_{g}$ orbital
degree of
freedom.\cite{K.Hirota98,I.Solovyev96,S.Ishihara96,D.N.Argyriou,Moritomo98}  It
is thus necessary to study the hole concentration dependence of structure in
detail and compare with the magnetic phase diagram we have established in the
present study.

%\newpage

\end{document}